\documentclass[sigconf, 10pt, nonacm]{acmart}
\usepackage[utf8]{inputenc}
\usepackage{outlines}
\usepackage{cancel}
\usepackage[compact]{titlesec}
\usepackage{xspace}
\usepackage{paralist}
\usepackage{caption}
\usepackage{xcolor}
\usepackage{booktabs}
\usepackage[subtle]{savetrees}
\usepackage{graphicx}
\usepackage[capitalise]{cleveref}
\usepackage{subcaption}

\usepackage{hyperref}
\hypersetup{colorlinks,linkcolor={blue},citecolor={blue},urlcolor={red}}

\newcommand{\weakcoherence}{federated coherence\xspace}

\newcommand{\WeakCoherence}{Federated Coherence\xspace}

\def\t{\textit}

\newcommand{\topheading}[1]{\noindent\textbf{#1}}
\newcommand{\heading}[1]{\vspace{4pt}\noindent\textbf{#1}}

\newcommand{\eat}[1]{}

\usepackage[normalem]{ulem}
\newcommand{\allnotes}[1]{}
\renewcommand{\allnotes}[1]{#1}

\newtheorem{definition}{Definition}

\newcommand{\notepanda}[1]{\allnotes{\textcolor{cyan}{[Panda: #1]}}}

\newcommand{\notemarcos}[1]{\allnotes{\textcolor{teal}{[Marcos: #1]}}}
\newcommand{\noteamaro}[1]{\allnotes{\textcolor{purple}{[Amaro: #1]}}}

\newcommand{\notevincent}[1]{\allnotes{{\color{blue}[Vincent: #1]}}}

\begin{document}

%\title[Getting Consistency from Incoherent Disaggregated Memory]{Getting Consistency from Incoherent\texorpdfstring{\\}{ }Disaggregated Memory}
\title{The Dawn of Disaggregation and the Coherence Conundrum: A Call for Federated Coherence}

\author{Jaewan Hong}
\affiliation{%
  \institution{UC Berkeley}
  \country{} % Leave empty
}
\author{Marcos K. Aguilera}
\affiliation{%
  \institution{VMware Research}
  \country{} % Leave empty
}
\author{Emmanuel Amaro}
\affiliation{%
  \institution{Microsoft}
  \country{} % Leave empty
}
\author{Vincent Liu}
\affiliation{%
  \institution{University of Pennsylvania}
  \country{} % Leave empty
}
\author{Aurojit Panda}
\affiliation{%
  \institution{NYU}
  \country{} % Leave empty
}
\author{Ion Stoica}
\affiliation{%
  \institution{UC Berkeley}
  \country{} % Leave empty
}

\maketitle

\section{Abstract}

%{\bf Abstract.}
Disaggregated memory is an upcoming data center technology that
will allow nodes (servers) to share data efficiently.
Sharing data creates a debate on the level
  of cache coherence the system should provide.
While current proposals aim to provide coherence for
  all or parts of the disaggregated memory,
we argue that this approach is problematic, 
  because of 
  scalability limitations and hardware complexity.
Instead, we propose and formally define \emph{federated coherence}, a model that provides coherence only within nodes,
  not across nodes.
Federated coherence can use current intra-node coherence
  provided by processors
without requiring expensive mechanisms for inter-node coherence.
Developers can use federated coherence with a few simple
  programming paradigms and a synchronization library.
We sketch some potential applications.

% Scaling memory capacity and enabling efficient data sharing are key drivers for the adoption of byte-addressable shared disaggregated memory in data centers.
% While current proposals aim to extend hardware cache coherence across disaggregated nodes, this approach encounters significant scalability limitations.
% This paper argues that the inherent cost of cross-node cache coherence becomes unsustainable as disaggregated memory deployments grow.
% We propose and formally define \emph{federated coherence}, a cache coherence model that pragmatically leverages existing intra-node coherence while eliminating the requirement and overhead of inter-node coherence.
% We demonstrate the viability of this approach by outlining effective programming paradigms for developing distributed applications that share in-memory state over disaggregated memory using federated coherence.

\section{Introduction}
\label{sec:intro}

Disaggregated memory is a recent trend in data center design, where additional memory
  is placed in external memory blades connected to nodes (servers) via a fast fabric.
Disaggregated memory is gaining traction due to the industry-wide adoption of a new standard, the Compute eXpress Link (CXL).
The latest CXL specification allows disaggregated memory to be used as a byte-addressable shared memory across nodes,
  providing
  a fast means of sharing data without
   incurring network overheads (serialization, deserialization, multiple copies).
This capability is attractive for modern distributed systems that process big
  data in memory~\cite{stephanie_ownership, exoshuffle, ramcloud, spark, rdds, graphx}.

The conventional wisdom is that shared memory should be cache coherent; otherwise it becomes intractable to program
  (\S\ref{sec:background}).
Thus, CXL aims to support cache coherence using derivatives of well-established coherence protocols, such as MESI~\cite{mesiTextBook}.
We argue that cache coherence is unsuitable for disaggregated memory (\S\ref{sec:issues}), for several reasons.
First, coherence protocols do not scale with the number of caches~\cite{MHS2012,scaleNUMA}, due to the coherence traffic required between them.
Even a small disaggregated memory deployment comprising 10 nodes with 100 cores each would have 1000 caches. 
%Yet, even a small disaggregated memory system can have thousands of caches
 % (e.g., in a system with 10 nodes and 100 cores per node).
Second, these protocols do not scale with memory size because of the space overhead of metadata.
Yet,
   disaggregated memory is targeted at large memories (PBs and beyond).
Third, implementing coherence across nodes introduces prohibitive hardware complexity.
%, as
%processors must handle asynchronous cross-node invalidation requests.
%  which are already challenging in single-node systems.
  %that are tightly coupled.
%coherence protocols are already complex inside
%  a tightly coupled server, but running such protocols across nodes makes things even more complex.
%For example, processors must be designed to respond to arbitrary asynchronous external requests for invalidation.
  
On the other hand, providing no cache coherence is unattractive from a developer's perspective.
Such systems have historically 
   failed due to their complexity and poor usability (\S\ref{sec:issues}).
%: without coherence, a fresh cache may be overwritten at any time with the stale contents of another cache.
This raises an important question: what level of coherence is appropriate for distributed applications to leverage shared disaggregated memory?

We answer the question by proposing and formally defining a coherence model for disaggregated memory, called \emph{\weakcoherence} (\S\ref{sec:newcoherence}), which presents unified address spaces to distributed programs.
Intuitively, \weakcoherence provides coherence only between caches in the same node.
Thus, coherence protocols only execute within nodes, which solves the scalability problems.
%  as the coherence protocol run only within nodes, there is no concern as we grow the system.
It also reduces hardware complexity, since there is no need for global coherence communication---in fact,
  as  current processors in data centers (in the x86 family) are all cache coherent,
  implementing \weakcoherence is straightforward (\S\ref{sec:newcoherence}).
%\notejae{What about this? We propose a new cache coherence model called \emph{\weakcoherence} (\S\ref{sec:newcoherence}). \weakcoherence strikes a balance between full coherence and no coherence by providing coherence only within individual nodes, while eliminating coherence across nodes. This design scales effectively, as intra-node coherence protocols remain lightweight and efficient, and cross-node communication is avoided. Furthermore, federated coherence imposes minimal hardware complexity, as modern processors (e.g., x86) already support cache coherence within a node, making implementation straightforward.}
%\noteamaro{Concerned with this para, and first sentence of next. This type of coherence segregation has existed before, so when we say ''new'', it might be questionable. For example, if a host has a SmartNIC, the host and the SmartNIC have disjoint coherence federations. A few ideas on how to distinguish: (1) federated coherence exposes a unified byte-addressable address space to user; (2) we formalize it. Thoughts?}
%\notepanda{I don't entirely understand the concern: in your example the SmartNIC and host have different address spaces, no? Or are you imagining some sort of unified address space, but different coherent domain? I agree that within a single server there might be many coherence domains, but I think they usually correspond to different address spaces?}

Fundamentally, \weakcoherence is a coherence model situated between full global coherence and no coherence.
A natural question for a weaker coherence model is how do developers use it to write distributed applications.
In \S\ref{sec:casestudy}, we present several paradigms for distributed programming that build on the concepts of node ownership, immutability, and versioning. We then sketch how these paradigms can be employed in microservices, pub/sub and immutable object store systems.

Under node ownership, each data item in disaggregated memory is
assigned to an owner node (which can change over time), so only threads within that node can coherently access the data item.
Thus, an owner node can use familiar synchronization primitives such as atomics and locks.
We also propose simple mechanisms to reassign ownership. For example, this works well in data pipelines where each stage of the pipeline can execute on separate nodes.
%For example, this paradigm works well in applications written as data pipelines, where each stage of the pipeline runs in a node.
Immutability ensures that, once a data item is created, the node that created the item flushes all its caches, ensuring global data item visibility.
%With versioning, we augment data items with a monotonic counter that gets attached to pointers to the data item, so that a thread
%  receiving the pointer can verify that it is accessing an up-to-date version.
  %(if not, it clears its cache).
Versioning augments data items with a monotonic counter, allowing threads to verify they have access to the most up-to-date version.
Lastly, we outline synchronization primitives tailored for \weakcoherence, including
  locks, semaphores, and other common primitives, so that threads in different nodes can coordinate without requiring global coherence.
Overall, we believe these paradigms offer developers a practical and scalable way to harness shared disaggregated memory under a weaker coherence model.
%This library can be built using algorithms that work with non-coherent memory.

%Using these ideas, we sketch a few disaggregated memory applications that can benefit from \weakcoherence: microservice-based systems,
%  a publish-subscribe system, and an immutable object store. 

We stand at an architectural inflection point in the design of disaggregated systems where we have realized that fully coherent
memory is not viable, so we need alternatives. The choice of coherence affects all aspects of disaggregated systems:
hardware architecture, programming model, software
design, and cluster management. Thus, it is critical for the broad community to discuss the proper coherence model and its trade-offs now. We wrote this paper to seed that conversation.
\section{Background}
\label{sec:background}

\subsection{Disaggregated Memory}
Disaggregated memory is memory that is physically segregated from compute \emph{nodes} (data center servers),
  residing
%\footnote{Nodes refer to data center servers, which typically have dozens to hundreds of processors in one or more CPU sockets, some local memory separate from the disaggregated memory, and IO devices.},
  in one or more memory blades reachable via a fast fabric.
We focus on byte-addressable disaggregated memory, which is directly accessible by processors in nodes via load and store instructions, rather than remote memory accessed using mechanisms such as RDMA or page faults.

Disaggregated memory has a long history~\cite{LCMRRW2009};
however, its recent surging interest is due to emerging commercial hardware support, including CCIX~\cite{ccix}, Gen-Z~\cite{genz}, OpenCAPI~\cite{opencapi}, and more recently CXL~\cite{cxl}.
CXL is now the dominant approach as it has gained broad support from a wide range of vendors and customers.
Early disaggregated memory deployments have 8--16 nodes~\cite{pond}; the future will see CXL v3 switches~\cite{cxl} that support 20--80 nodes and beyond.
%become an industry standard with 

A key feature of this new hardware is the ability to share data across nodes, serving as a shared memory.
There is a longstanding debate about whether message passing or shared memory is the best way to build distributed systems.
We do not intend to settle this debate; instead, we simply investigate the opportunity of disaggregated memory to \emph{scale up} applications by running threads across nodes that share data in disaggregated memory, and we explore the constraints of the hardware-software design space.

% In this paper, we focus on distributed systems that use shared disaggregated memory\footnote{In this paper,
%   the term (shared) disaggregated memory refers to
%   memory addressable by 
%   load and store instructions, as opposed to mechanisms such as
%   RDMA or page faults (which are less concerned
%   about cache coherence).} for communication. There is a longstanding debate about whether message passing or shared memory is the best way to build scalable systems. We do not take a position on this debate; instead, we note that existing hardware has made it feasible to build shared memory systems using off-the shelf parts. At the same time, the increasing memory demands of modern workloads, and a slowdown in the
%   improvement of local memory capacity has led to increased adoption of disaggregated memory systems. These systems allow multiple servers to combine their compute power to
%   process data placed on large pools of shared memory, and could 
%  provide the computational capacity of an expensive scale-up supercomputer~\cite{nvidia_dgx_gh200} using commodity hardware.
% While early disaggregated memory deployments have 8--16 sockets~\cite{pond}, we envision future rack scale
% system with 20--80 servers using CXL v3 switches~\cite{cxl}.

\subsection{Issues with Cache Coherence}
\label{subsec:unscalable}
\label{sec:issues}

Conventional wisdom says that shared memory should be cache coherent~\cite{coherencetextbook}, meaning that CPU caches must be kept consistent with each other, i.e., the same address cannot be validly cached at different CPUs with different data.

Cache coherence is ubiquitous in modern systems (e.g., every x86 server is cache coherent) because of its conceptually simple programming model.
Consequently, the CXL specification adds support for sharing memory with coherence in version 3.0, primarily using two mechanisms:
snoop filters and back invalidation~\cite{backInvalSnoopCaches}.
Roughly speaking, a snoop filter within the CXL memory device tracks nodes that are potentially caching a given memory location. 
Upon detecting an update, the device initiates a back invalidate request to those nodes, compelling cache invalidation.
%VL: removing for space
%This resembles directory-based coherence~\cite{dash}, albeit with reduced precision (node-level tracking, not individual caches) and coarse granularity (coalescing multiple cache lines).

\begin{figure}[t]
    \centering
    \includegraphics[width=0.7\columnwidth]{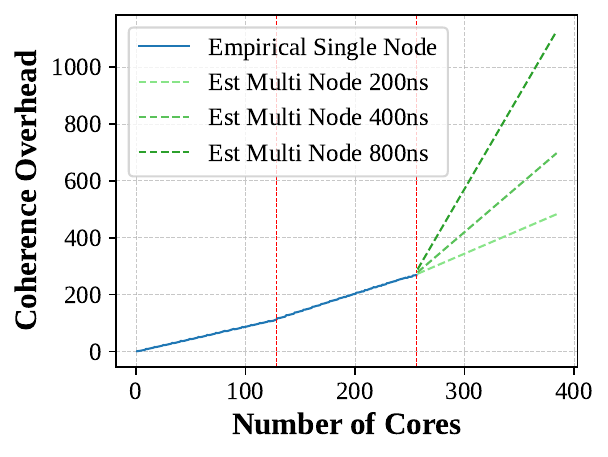}
    \caption{Overhead of cache coherence vs. core count, where each core contains a pinned thread that repeatedly increments a shared global variable atomically.
    We measure the aggregate rate of increments per second---coherence overhead is the ratio between the rate in a non-cache coherent and cache-coherent system.
    Non-cache coherence is emulated by having each thread increment a different variable. 
    The solid line plots measured results, while the others are extrapolated using varying latency to disaggregated memory.
Red vertical lines mark NUMA node boundaries.
    }
    \label{fig:extrapolate}
\end{figure}

We argue that cache coherence will not work with disaggregated memory, whether it is provided by
  the current CXL proposal or other means (cache snooping, directories, and variants
  thereof), for three reasons:
  
%\begin{enumerate}
%\item 

(1) \textit{Cache coherence fails to scale with the number of disaggregated participants.}
The underlying fabric on which a coherence protocol would run has higher latency and lower bandwidth than local NUMA links, partly because the wires between disaggregated components are inherently longer.
This translates to slower coherence protocol execution.
%Consequently, the cache coherence protocol would be slower than it would within a host, or a NUMA node.
This is on top of the classic scaling issues with snooping; while directories
  fare better, they still cannot scale to the hundreds or thousands of caches in a
  disaggregated memory system (\cref{fig:extrapolate}).
%The overhead increases nearly linearly with the number of cores, reaching 244 at 256 cores.

%\item 
(2) \textit{Cache coherence fails to scale with memory size.}
Snoop-based mechanisms are known to incur unsustainable bandwidth overheads for large memory, while directory-based mechanisms and their variants consume too much space---including CXL's snoop filters~\cite{danielberger,AMDCXL}.
Prior work \cite{MHS2012} has argued that cache coherence scales within a single SoC by providing precise sharer tracking, thus keeping overheads largely independent of core count via minimized invalidations and acknowledgments. However, maintaining this precision across hundreds or thousands of participants would incur additional storage costs, rendering the approach infeasible.
%\notevincent{is this memory space or die space?}
  %cannot handle the volume of traffic of a large memory, while directory-based mechanisms and
  %\noteamaro{this last part of ''space'' fits better with the next reason; i think we might need to merge both points}
  
%\item 
(3) \textit{Cache coherence introduces prohibitive hardware complexity.}
Cache coherence protocols are among the most complex logic in a processor~\cite{mesi}, and cross-node coherence is even more complex.
In fact, despite significant effort, the CXL protocol for coherence has been found to be underspecified~\cite{cxlformalising}.
This complexity, particularly the sophisticated lookup/update logic of CXL's snoop filters, further grows exponentially with the number of coherent agents~\cite{AMDCXL} and serves as a die-area bottleneck~\cite{danielberger}.
In addition, CXL's back invalidate add further complexity
  as it requires that processors 
  handle asynchronous external requests to invalidate cache 
  lines.
%we have yet to see a processor with that capability.
  
  %and while we have managed to provide coherence within a single node, doing so across nodes is even harder.
%In particular, (1) the CXL protocol for coherence is ambiguous and not fully specified~\cite{cxlformalising};
  
An alternative approach is to forgo 
cache coherence, an idea explored in both research~\cite{runnemede, incoherent_hw}
  and industry~\cite{ps3_cpu, arm_oldcpu}.
%This direction has also been explored by industry, for example, the Cell processor of the Playstation 3, and old ARM CPUs were not fully cache coherent~\cite{ps3_cpu, arm_oldcpu}.
This approach scales well and avoids the hardware complexity of coherence.
However, a non-coherent system adds significant complexity to applications and prevents multi-threaded applications from efficiently sharing data.
Thus, none of those prior approaches have been widely adopted.

%\heading{Partial/weaker coherence.}
Recent work on CXL proposes a partitioned approach, with a small cache-coherent memory region alongside an  incoherent region~\cite{pasha,danielberger,AMDCXL}.
%
%on CXL proposes the idea of partitioning the disaggregated memory into two parts:
  %a small region that is cache coherent and a larger region that is not.
%Then, critical parts of an application can run in the coherent region.
This approach addresses CXL's memory-size scaling issues, 
%because it reduces the overhead of CXL snoop filters as only the small coherent region needs to be covered.
but not the other problems (scaling with
the number of participants and hardware complexity):
the overheads of \cref{fig:extrapolate} still hold for
small regions, and hardware must still provide full support for coherence of the small region. 

Cosh~\cite{cosh} proposes the concept of coherence islands,
  which inspires federated coherence, but in a different
  context: Cosh defines OS-level I/O abstractions for processes to share aggregates (byte buffers) over single-node coherent and incoherent shared memory,
  rather than shared memory regions that
  threads can access via loads and stores.

Meanwhile, early prior work has explored 
  the idea of weakening coherence in multiprocessor
  systems (e.g., \cite{dash}) and distributed shared memory
  (e.g., \cite{ivy}) to improve performance by deferring cache invalidations.
%\notejae{dumped related work of weaker consistency and coherence. Need compactin}
%We are not the first to propose new levels of coherence.
There is also a rich history of research exploring
weaker \emph{consistency} models for improving performance,
 such as Release Consistency~\cite{rc},
 Entry Consistency~\cite{EC}, and
 Lazy Release Consistency~\cite{LRC}.
We observe that coherence and consistency are orthogonal
considerations~\cite{SWD2011}: the former is concerned about the behavior of caches, while the latter is concerned with 
reordering of operations by processors. 

\section{\WeakCoherence}
\label{sec:newcoherence}

An effective coherence model for shared disaggregated memory must balance two factors $(a)$ what can be implemented efficiently in a shared-memory disaggregated system and $(b)$ what is required by applications that run on it.
Thus, our target applications share three key characteristics: (1) they have a large working set sizes, 
    (2) they rely on a small number of threads actively sharing and writing to the same memory (or operate in read-only mode) and thus benefit from fine-grained synchronization;
    and (3) they eventually need coarse-grained coordination (e.g., a reduce stage in MapReduce).

%The novelty of our proposal is to identify a level of coherence that is well suited for disaggregated memory.
%\notevincent{key idea: design coherence with the hardware in mind -- locality}
%The rationale for this form of coherence is twofold.

% Marcos' most recent text
%For the second, we expect that in disaggregated memory applications, threads within a node are more tightly coupled than threads across nodes, i.e., threads in the same node are more likely to share data, synchronize, and hence benefit from coherence.
%We examine this assumption through a few case studies (\S\ref{sec:sampleapps}).

%Of course, this expectation reflects an assumption about how disaggregated
%  memory applications are constructed, which we later examine in a few case studies (\S\ref{sec:casestudy}).

%\heading{Basic concept and rationale.}
Our proposal, \WeakCoherence, provides coherence exclusively based on physical locality.
Practically, this means that within a node, a location is either in shared mode in all caches or exclusive mode in at most one cache.
This property does not hold across nodes;
for instance, a location may be in exclusive mode in different caches at different nodes.

\WeakCoherence is easy to implement.
In fact, some current CXL disaggregated memory deployments already offer \WeakCoherence, albeit unintentionally:
shared regions deem\-ed non-coherent in CXL (where the snoop filter and back invalidations are missing), effectively provide \WeakCoherence, since nodes connected to the disaggregated memory provide internal coherence with no global coherence mechanisms.
Thus, applications get a free lunch if they are designed to exploit \WeakCoherence rather than no coherence.

\WeakCoherence addresses the three issues with cache coherence (\S\ref{subsec:unscalable}).
First, it scales well as we add more nodes because there is no coherence traffic between nodes.
Second, it scales well as we add more disaggregated memory because its implementation incurs no space overheads for snoop filters or other coherence mechanisms.
Third, it avoids any hardware complexity because it requires no support from CXL devices; it leverages existing coherence mechanisms.

Finally, \WeakCoherence also addresses the main issue with no coherence: high software complexity.
  As we later explain (\S\ref{sec:casestudy}), there are a few simple paradigms for developers 
  to use \WeakCoherence
  efficiently and correctly.
These paradigms are applicable to widely different types of applications (\S\ref{subsec:sampleapps}),
  and they can also be used to provide drop-in replacements for traditional synchronization and
  data structure libraries.
They also enable developers to leverage federated coherence for large, hierarchical workloads with semantics that
  fit the data structure without prohibitive overhead or complexity.

\heading{Definition.}
We define \weakcoherence following two common practices~\cite{SWD2011}.
First, coherence is defined separately from the memory model,
  by considering the order of operations issued by processors to the cache
  controller rather than the program order.
  %(which may allow instruction re-ordering and other
  %quirks)---thus, the coherence definitions are with respect to the
  %order of operations issued by the processors
  %to the cache or memory controller, rather than
  %the program order.
% Second, the definitions cover only load and store operations
%   not read-modify-write operations (compare-and-swap, fetch-and-add, etc).
Second, coherence is defined for a single fixed memory location rather than the entirety of memory.
We start with the standard 
  cache coherence definition~\cite{SWD2011},
  where memory operations are 
  $\t{write}_p(v)$ and $\t{read}_p(v)$, 
  representing a store and load of $v$ 
  by processor $p$ at a fixed memory location.

\begin{definition}
\label{def:coherence}
  A memory system satisfies \emph{cache coherence} if, for every execution and each memory location $\ell$, there is a total order $\mathcal{O}$ of operations on $\ell$ in the execution, such that:

  \begin{enumerate}
  \item $\mathcal{O}$ is consistent with the order of operations on $\ell$ issued by each processor $p$.%\notepanda{Order of operations makes this sound like we are assuming SeqCst? Maybe issued is the problem?}

  \item In $\mathcal{O}$, a read $r$ returns the value of the last preceding write.
  %that precedes $r$.
  \end{enumerate}
\end{definition}

\noindent For convenience, we assume that the execution
starts with $\t{write}_{p_0}(v_0)$, representing
the initialization value $v_0$ of $\ell$ by a
hypothetical processor $p_0$.
Thus, Definition~\ref{def:coherence} states that operations on a memory location can be serialized,
such that a read always
returns the value of the latest write.
%(where the
%serialization may differ from location to location).

For weaker forms of coherence, we extend processor operations
  to include $\t{flush}_p$, 
  representing a flush of the cache of processor $p$ to memory.
%
  % \noteamaro{Can we unify the definition of flush to always flush from a shared cache? I believe this modified condition doesn't require us to make the flush here only apply to processor p's cache. This would allow us to have a single definition of flush}
  % \notemarcos{if we use a single shared cache in this definition, we get full coherence because caches cannot diverge}
%  
We now define a system with
  a weak form of coherence where processors
  can explicitly flush their caches.
This definition is obtained by modifying condition (2) in Definition~\ref{def:coherence} as follows:

{\it
  \begin{enumerate}
  \item[(2)] In $\mathcal{O}$, a read $r$ returns a
  value picked as follows:

  \begin{enumerate}
      \item If the last operation by $p$ preceding $r$ is $\t{write}_p(v)$ or $\t{read}_p(v)$, pick value $v$.

      \item If the last operation by $p$ preceding $r$ is
      $\t{flush}_p$,
\eat{
      $r$ returns the value $v$ from the last $\t{write}_q(v)$
      preceding the most recent $\t{flush}_q$ by any processor $q$.\noteamaro{check--original eaten below}
      \notemarcos{the notion of ``most recent'' isn't
                 clear.}
}
      find the process $q$ with the
      last $\t{flush}_q$ preceding $r$
      %(this could be $q=p$)
      and pick the value $v$ from the
      last $\t{write}_q(v)$ preceding the $\t{flush}_q$.

  \end{enumerate}

  \end{enumerate}
}

Intuitively, this condition says that a read should
 return either the value in the local cache or the
last value that was flushed to memory, depending on whether the reading processor recently flushed its cache.
For convenience, we assume the execution
starts with $\t{write}_{p_0}(v_0)$ followed by $\t{flush}_p$ for every processor $p$.
%\notevincent{the initialization probably needs updating here (fake flush)?}

\eat{
  We next give a definition of \weakcoherence
  where
  all flushes are explicitly issued by the
  program (no cache eviction by the system).
  We obtain this definition by replacing condition (2) of Definition~\ref{def:coherence} with the following:

  {\it
  \begin{enumerate}
  \item[(2)] In $\mathcal{O}$, a read $r$ returns one of the following:

  \begin{enumerate}
      \item If the last operation by some process $p'\in\t{node}(p)$ preceding $r$ is $\t{write}_{p'}(v)$ or $\t{read}_{p'}(v)$, pick the value $v$.

      \item If the last operation by some process $p' \in \t{node}(p)$ preceding $r$ is
      $\t{flush}_{p'}$, find the process $q$ with the
      last $\t{flush}_q$ preceding $r$ and pick the value $v$ from the
      last $\t{write}_{q'}(v)$ preceding the $\t{flush}_q$ with $q'\in\t{node}(q)$.
  \end{enumerate}

  \end{enumerate}
}
}

We now 
  define \weakcoherence
  by modifying the previous definition in two ways.
First, we extend the cache to be shared by all the processors in a node,
  where the intended behavior of a $\t{flush}_p$ operation is to flush this shared cache.
%\noteamaro{should we explicitly say this was not true before?}
%MKA: done, we now say earlier that flush_p represents a flush
%of the cache of processor p.
Second, we allow the total order $\mathcal{O}$
  to include additional flush operations not
  present in the execution,
  representing cache evictions by the hardware.
  %\noteamaro{what system?}.

\begin{definition}
\label{def:weakcoherence}
  A memory system satisfies \emph{\weakcoherence} if, for every execution
  and each memory location $\ell$, there is   
  a total order $\mathcal{O}$ that includes all
  operations on $\ell$ in the execution and
  possibly additional $\t{flush}$ operations, such that:

  \begin{enumerate}
  \item $\mathcal{O}$ is consistent with the
      order of operations on $\ell$ issued by each processor $p$.

  \item In $\mathcal{O}$, a read $r$ returns a value picked as follows:

  \begin{enumerate}
      \item If the last operation by any processor $p'\in\t{node}(p)$ preceding $r$ is $\t{write}_{p'}(v)$ or $\t{read}_{p'}(v)$, pick $v$.

      \item If the last operation by any processor $p' \in \t{node}(p)$ preceding $r$ is
      $\t{flush}_{p'}$,
\eat{
      then $r$ returns the value $v$ from the last $\t{write}_{q'}(v)$ preceding the most recent $\t{flush}_q'$ by any processor $q'$, where $q' \in \t{node}(p)$.\noteamaro{check--original eaten}
}
      find the process $q$ with the
      last $\t{flush}_q$ preceding $r$ and pick the value $v$ from the
      last $\t{write}_{q'}(v)$ preceding the $\t{flush}_q$ with $q'\in\t{node}(q)$.
      \eat{\notepanda{This one is a bit strange. What happens when $p, q\in \t{node}(p)$?}
      \notemarcos{$p$ is always in $\t{node}(p)$, so nothing different
      happens. If $q \in \t{node}(p)$, this corresponds to
      the node flushing its cache then reading again from memory.}
      \notepanda{Also what happens if $\t{node}(p)$ does not have the location cached,
      and also does not issue a flush?}
      \notemarcos{Impossible. If the location is not cached, it's
      because a flush happened by the program or due to
      an eviction (that's why $\mathcal{O}$ may include additional
      flush operations)}
      \notepanda{Also, if $\t{node}(p) = \t{node}(q)$, and there is a $\t{write}$ from $\t{node}(q)$ between $\t{flush}_q$ and $r$, do we loose the write?}
      \notemarcos{Impossible. That falls in case (a) above not (b).}}
  \end{enumerate}

  \end{enumerate}
\end{definition}

\noindent Here $\t{node}(p)$ denotes the set with the processors
in the same node as $p$, and
%\noteamaro{Can we add an intuitive summary of def 2?}
$\t{flush}_p$ is an operation by $p$ to
  flush the cache shared by processes
  in $\t{node}(p)$.
Intuitively, Definition~\ref{def:weakcoherence} says that in the serialized order, reads should return either the value in the cache shared by the node, or the last value flushed to memory from one of the shared caches, depending
on whether the reader's shared cache has been flushed.
%by a process in the reader's node.

% The above definition covers only loads and stores but it is
%   easy to extend it to read-modify-write
% operations (\eg compare-and-swap, fetch-and-add).
% \notemarcos{does it make sense for the processor to issue atomic
% operations to the cache or memory controller?}

\heading{Anomalies.}
\label{sec:anomalies}
With \weakcoherence, two types of anomalies are possible:
  cross-node stale reads and cross-node broken atomicity.
Cross-node stale reads happen when a read 
  runs at a node different from the one that
  issued the last preceding
  write, causing the read to miss the write and
  return a stale value.
Cross-node broken atomicity happens when
  atomic operations run at two different nodes
  causing them to break their atomic behavior (e.g.,
  two compare-and-swap atomics may succeed when only one was supposed to, or two fetch-and-increment atomics may increment a counter by only one).
These anomalies break existing concurrent data
  structures (e.g., lock-free queues), synchronization primitives
  (e.g., locks, semaphores), and consistency guarantees
  expected by most developers 
  (e.g., release consistency).
  %\footnote{Release consistency ensures that modifications done while a lock is held
  %become visible to subsequent lock holders.}

However, with \weakcoherence, such anomalies happen
  only across nodes.
If a read runs on the same
  node as the last preceding write, the read
  returns the correct value.
If two atomic operations run on the
  same node, they preserve their atomic guarantees.
These guarantees are the basis for developers
 to use under \weakcoherence, which we discuss next.

\eat{
\subsection{Implementation}

\notevincent{Since we're over on space, this subsection is a candidate for deletion.}\notepanda{I agree, deleting this if necessary seems fine.}
While a complete design for hardware that provides disaggregated memory with \weakcoherence is well beyond the scope of this paper, here we briefly sketch a possible approach.
The basic idea is that each node runs its internal hardware-based
  coherence protocol as provided in modern processors.
We connect all nodes to the disaggregated memory without additional coherence mechanisms so that when a node flushes its cache to memory, it immediately overwrites the previous contents without coordinating with other
  caches.
Note this implementation is efficient because the coherence traffic is constrained within each node.
% \notevincent{This section is a bit sparse.  Is there anything else we can add? If no time, it's fine to keep as is.}
}

\eat{
We focus on the CXL v3 hardware
  as the leading contender to realize shared
  disaggregated memory.
  \notepanda{Is there a particular thing we do
  in the proposal below that relies on anything from CXLv3?
  Especially given we disable the coherence engine?}
As we explained before, CXL v3 provides (full)
  cache coherence
  through back-invalidate messages from the disaggregated
  memory controller (home agent).
Basically, we obtain a system with \weakcoherence by
  omitting such messages and the snoop filter mechanism that
  supports them.
\notemarcos{Emmanuel: isn't there a way to disable
cache coherence in CXL (e.g., by definition regions
that are not coherent)? How did that work?}\noteamaro{The spec says regions can have software-defined coherence. I think the way it worked in Niagara is, they just exposed the regions to multiple hosts and that's it}
}

\section{Using Federated Coherence}
\label{sec:casestudy}

We now explain how to use Federated Coherence to develop applications.
We start with general paradigms and then cover some sample applications.

\subsection{General Paradigms}
\label{sec:paradigms}

\topheading{Node ownership.}
This paradigm ensures
at most one node accesses a data item at a time, by assigning a node
  to be the owner of the item.
Then, given the guarantees of Federated Coherence, threads within that node can use
  traditional techniques for synchronization and shared access designed for fully
  coherent memory, such as atomic operations, locks, etc.
A data item is an application-defined unit, potentially spanning multiple cache lines (e.g., it could be a struct, a fixed-size buffer,
  a variable-length list, a data structure, etc).
To change owners, applications can use different mechanisms depending on the frequency required.
For infrequent changes (e.g., once every few milliseconds), applications can use
  message-passing communication between nodes (e.g., over TCP): a thread in the owner node sends a message to the new
  owner.
For more frequent handoffs, a dedicated location in disaggregated memory can store the current owner's ID.
The current owner updates this ID and flushes the cache line.
Potential new owners flush and read this location to check for ownership.

\heading{Pipeline processing.}
Data processing applications can be organized as a pipeline where data items in disaggregated memory
  are processed in stages, each stage served by the threads of a node.
Then, using federated coherence, each stage sees the data items coherently across its threads.
Stages are connected by tasks queues (provided by a synchronization library, see below), 
  and when an item moves from one stage to another
  we flush its cache lines at the source and destination nodes.

% Each worker pool resides within a single Federated Coherence domain and processes tasks in a pipeline-like architecture.
% Unlike traditional event-driven architectures, where each stage enqueues jobs into the subsequent stage’s job queue (often located in globally shared memory), Federated Coherence enables a more localized approach. 

% Each stage of the pipeline maintains its own job queue in its local memory region in the disaggregated memory. When a stage produces a job, it writes the job entry to its designated memory region and flushes the cache line to ensure visibility to subsequent stages.
% The downstream stage reads jobs from the upstream stage’s job queue by directly accessing the corresponding memory region. Since Federated Coherence guarantees coherence within a node, worker threads in the same node can safely and efficiently process the job queue without additional synchronization mechanisms.
% Upon completing a job, the worker pool writes a completion record to a job completion queue in its local memory. This allows the upstream stage (or a dedicated management thread) to mark the job as completed and remove it from the job queue.

% \notejae{Concise version}
% Within a Federated Coherence domain, each worker pool operates as a pipeline stage with a local job queue. Unlike traditional globally-queued event processing, producers write jobs to their local memory and flush. Consumers directly read from the upstream queue. Intra-node coherence ensures safe local processing. Completion is recorded locally for eventual removal.

\heading{Immutable items.}
A node may produce immutable data consumed by other nodes (e.g.,
  a buffer read from a file, or items in an immutable key-value store~\cite{ray_2018}).
The producer node simply flushes its cache once
  the immutable item is finalized, and the other nodes
  can read the item's latest version by using non-temporal
  reads or flushing their caches prior to reading\footnote{Note
  that this does not cause write backs since data is immutable.}
A common case of immutable items is parameters of 
  RPCs, where both parameters and results are treated as immutable. 
Garbage collection of immutable items can be managed using a shared \emph{freed} flag in disaggregated memory.
The node releasing the item sets the flag and flushes its cache,
while a garbage collection thread periodically flushes and checks the flag.

\heading{Synchronization across nodes.}
We implement a software library for Federated
  Coherence that provides implementations of locks, semaphores, queues, and other
  synchronization primitives, so that threads in different nodes
  can coordinate.
These implementations can use algorithms
  that work with non-coherent memory, such as Lamport's Bakery Algorithm~\cite{bakery}, modified Peterson's lock~\cite{petersons}, and token-based locks~\cite{wagner2000token}.
This library is complex to implement, but it needs to be done only once by an expert and can be reused multiple times.

\heading{Version numbers.}
A data item can be associated with a version number so that threads in different nodes can refer to the correct version---for example, when threads
   pass a pointer to the item, it attaches the intended version number.

%\vspace{-0.1cm}
\subsection{Sample Applications}
\label{subsec:sampleapps}

We can build a broad range of applications that span multiple nodes for scalability,
  using disaggregated memory with Federated Coherence, as we now illustrate.

\heading{Microservice-based systems.} Each node runs a different microservice,
  and nodes communicate with RPCs using disaggregated memory to efficiently
  pass parameters~\cite{dmrpc,hydrarpc}.
As mentioned above, parameter passing in RPCs can be done easily with
Federated Coherence.
To synchronize the execution of microservices (e.g., active the execution of a
  remote procedure), we can use the Federated Coherence synchronization library
  indicated above.

\heading{Publish-subscribe systems.}
We can implement an efficient publish-subscribe system by storing its shared
  log in disaggregated memory.
We partition the shared log into multiple large circular buffers, with each partition assigned to a node.
A node is allowed to write to its partition and read the partition of other nodes.
Publishers append messages to their log partitions, which become immutable items,
    and subscribers on other nodes read the items.
Version numbers on log entries allow for reuse of space in the circular buffers.
Publishers can scalably write to their buffer in multi-threaded with federated coherence.
Publishers and subscribers synchronize using the Federated Coherence synchronization library.

\heading{Immutable object stores.}
Common in distributed system frameworks like Ray~\cite{ray_2018}, these stores hold objects written by tasks or actors.
Once written, objects become immutable.
Frameworks then pass references to these objects, and tasks or actors retrieve them as needed.
%The immutability of the objects eliminates the need for strict cross-node coherence once the object is written, making Federated Coherence a suitable model.

\heading{Memory pooling system.}
With memory pooling, a region of disaggregated memory is allocated to extend the local memory of a VM.
Each allocated region is exclusively owned by a single VM's node, benefiting from the full intra-node coherence provided by federated coherence. 
Upon VM termination, the region is returned to the pool (node ownership transfer).
A control plane, implemented using message passing or the federated coherence synchronization library, manages the assignment of memory regions to nodes. This control plane is invoked infrequently during VM startup and shutdown.
\section{Call to Action}
\label{sec:calltoactions}

Disaggregated memory stands at a critical juncture: defining its cache coherence model.
Hardware vendors, software developers, and software researchers need
to get together to agree on a model that (1) hardware vendors can implement well, with good
performance and scalability, and (2) software developers can use with reasonable effort.
It is evident that neither full coherence nor complete incoherence are viable.

While we believe Federated Coherence is the right model,
what is more important is to get the discussion going, and do so
urgently while the hardware is still in its early stages.
In the process of developing this model, we need to define new
software benchmarks to measure the performance of end-to-end tasks
that relate to sharing data in disaggregated memory,
such as thread synchronization (e.g., semaphores, condition variables)
communication between threads (e.g., producer-consumer, broadcast).

Once we settle on a coherence model, additional research is needed to
(1) develop new paradigms, synchronization libraries, and data
  structure libraries for developers,
(2) exploit modern programming languages like Rust, which
natively support ownership to optimize
code for the weaker coherent model,
and (3) devise innovative techniques to find functional and performance
bugs.

\section{Conclusion}
\label{sec:conclusion}

Disaggregated memory can serve as a shared memory across nodes to scale up systems.
But we believe the community is moving in the wrong direction in its attempt
  to provide cache coherence for all or parts of disaggregated memory.
A better approach is federated coherence, which provides coherence only within nodes.
Some disaggregated memory systems already provide this form of coherence, without realizing it,
  so applications might as well use it.
For that, we have formally described the property and provided simple paradigms to use it.
Regardless of whether one believes this is the best approach, we need to get the discussion going.

% Federated Coherence is not a compromise---it is a deliberate and pragmatic balance between performance and complexity. 
% By focusing on locality, it addresses the shortcomings of fully non-coherent and fully coherent models, providing a scalable and practical solution for general-purpose disaggregated systems.
% Importantly, Federated Coherence serves as a solid foundation for future advancements.
% \notejae{For future work: Expand the coherence domain over a node boundary.}

% By adopting Federated Coherence, 
%     the community can accelerate the adoption of disaggregated memory, 
%     unlocking its transformative potential while avoiding the pitfalls of 
%     all-or-nothing coherence models. 
% It offers the best of both worlds: the simplicity and performance of local coherence,
%     combined with the scalability and efficiency 
%     required for the next generation of memory systems.

% We urge researchers, hardware designers, and software developers 
%     to seriously consider Federated Coherence as a viable and highly promising approach.
% It is time to move beyond the binary thinking of the past and embrace a 
%     coherence model that reflects the realities of 
%     modern workloads and hardware architectures. 
% Federated Coherence is the key to bridging the gap between performance, 
%     scalability, and practicality in disaggregated memory systems.

\balance

\bibliographystyle{plain}
\bibliography{bibliography}

\end{document}